\documentclass[format=sigconf, review=false, anonymous=false, screen, dvipsnames]{acmart}
\setcopyright{acmlicensed}
\copyrightyear{2018}
\acmYear{2018}
\acmDOI{XXXXXXX.XXXXXXX}

\acmConference[]{}{}{}
\acmYear{}
\copyrightyear{}
\acmPrice{}
\acmDOI{}
\acmISBN{}
\setcopyright{none}

\usepackage{booktabs} % For formal tables

\usepackage{exii-macros}

\usepackage{booktabs}

\definecolor{myyellow}{HTML}{F5E150}
\definecolor{mygreen}{HTML}{1D9521}
\definecolor{myblue}{HTML}{0C10E6}
\definecolor{myred}{HTML}{FF1414}
\definecolor{myblack}{HTML}{000000}
\definecolor{mypurple}{HTML}{8703DA}
\definecolor{mypink}{HTML}{FF1BE1}
\definecolor{mylblue}{HTML}{7D93EB}

\hidecomments
\hideoutline

\showrevisions{REVISIONGREEN}

\makeatletter
\g@addto@macro\normalsize{%
  \setlength\abovedisplayshortskip{-9pt}
  \setlength\belowdisplayshortskip{3pt}
}
\makeatother

\begin{document}

% this seems to help remove words going beyond the margin
\tolerance=400 

% The title should be short but descriptive, like a mini abstract. 
% If possible, come up with a catchy name for your project and use it as part of the title.
% can include a short version of the title for the running header (in square brackets)
\title[Exploring Visual Software Testing Output]{Exploring the Output of Software Testing Tools through a Visual Comparative Analysis}
% \titlenote{Produces the permission block, and
% \subtitle{Extended Abstract}
% \subtitlenote{The full version of the author's guide is available as
%   \texttt{acmart.pdf} document}


\author{Brandon Lit}
\orcid{0009-0002-9418-2695}
\authornote{All authors contributed equally to this research.}
\affiliation{%
  \institution{Cheriton School of Computer Science, University of Waterloo}
  \country{Waterloo, Ontario, Canada}
}
\email{bjlit@uwaterloo.ca}

\author{Anthony Maocheia-Ricci}
\orcid{0009-0002-3881-7166}
\authornotemark[1]
\affiliation{%
  \institution{Cheriton School of Computer Science, University of Waterloo}
  \country{Waterloo, Ontario, Canada}
}
\email{anthony.maocheia-ricci@uwaterloo.ca}

\author{Thomas Driscoll}
\orcid{0009-0002-7661-8811}
\authornotemark[1]
\affiliation{%
  \institution{Cheriton School of Computer Science, University of Waterloo}
  \country{Waterloo, Ontario, Canada}
}
\email{thomas.driscoll@uwaterloo.ca}

% If default list of authors is too long for headers.
\renewcommand{\shortauthors}{Lit et al.}

\begin{abstract}

Software testing is a fundamental process of software development, and prior work has shown that visualizations of test results support testers’ decision-making. However, Human-Computer Interaction research on software testing has yet to explore and understand the shared interface elements and patterns in visualization of testing outputs. To address this, we conducted a visual comparative analysis of the output of 50 software testing tools and harnesses (44 with CLI output, 6 with GUI output) across four popular programming languages. Our analysis reveals the common interface elements in software testing tools, how these tools display and visualize test results, as well as the specific make-up of the output. Our findings provide insight on how visual testing output is formatted and how colour is used across both CLI and GUI environments, identifying trends that can be applied by developers of testing tools.

% Prior work in human-computer interaction related to software testing has not explored building an understanding of these tools in terms of their shared interface elements and patterns in visualization.

% 30 word contribution statement (used for final version, but good to write now)

\end{abstract}



% keep keywords to one line in rendered paper, try to use big topics that aren't
% already in your title
% \keywords{interaction techniques, controlled experiments}

% % optional full width teaser figure
%   \includegraphics[width=\textwidth]{figures/testingoutputs.png}
%   \caption{Abstract wireframes of our 50 testing platform outputs used as our unit of analysis for visual comparison.}
%   \Description{This figure presents a mosaic of abstract wireframes representing the outputs of 50 different unit testing platforms. Each colored block represents a single testing platform. This mosaic serves as the unit of analysis for the subsequent visual comparison of these testing platforms.}

\maketitle

% --------------------------------
% BODY
% edit this file to insert your sections
% Exii Standard Section Index
% ================================

% sections are each in separate files

%!TEX root = paper.tex
\section{Introduction}

Software testing is a fundamental process of the software development pipeline. Much empirical research in software testing explores areas such as evaluation metrics~\cite{chenRevisitingRelationshipFault2021} or automated testing methods or frameworks (e.g., fuzz testing~\cite{kleesEvaluatingFuzzTesting2018, chenEnFuzzEnsembleFuzzing2019}, the use of Large Language Models~\cite{yangWhiteFoxWhiteBoxCompiler2024, xiaFuzz4AllUniversalFuzzing2024, ouMutatorsReloadedFuzzing2025}), but little work has explored the outputs of these testing tools and harnesses with respect to the end-user's interactions. While some software testing tools' output exists only as file structure (e.g., YARPGen~\cite{livinskiiRandomTestingCompilers2020}), other platforms include visual output through dashboard websites, or are dedicated GUI tools themselves. Studies in software testing show how processes benefit from having visualizations of test results ranging from a colour gradient to denote bug severity to full graphical user interface (GUI) dashboards~\cite{strandberg_information_2019}. Other qualitative studies in software engineering related to software testing point towards the usefulness of visualizations~\cite{ganChallengesStrategiesImpacts2025} or graphical outputs of the tests, their coverage, and properties of interest~\cite{pradoCognitiveSupportUnit2018}, which can serve as decision support tools for software engineers~\cite{strandbergDecisionMakingVisualizations2018}.  

Human-Computer Interaction (HCI) research is well-equipped to explore this space of end-user interactions with software testing tools. Existing HCI studies related to software testing focus primarily on accessibility~\cite{chiou_bagel_2023} and GUI testing tools or platforms~\cite{alaboudi_hypothesizer_2023, chen_improving_2020, liu_guided_2022, morgado_impact_2019}. However, building an understanding of these tools in terms of their shared interface elements and patterns in visualization has not yet been explored. As such, to further bridge the gap between HCI and software testing, we employ a {\it visual comparative analysis} methodology to explore the visual {\it output} of software testing tools and harnesses as a whole. Visual Methods are a family of qualitative methodologies primarily used in the social sciences to critically interpret visual materials such as fine artwork, maps, and digital images or photography~\cite{rose_visual_2023}. We adapt prior work using visual methods in HCI research~\cite{frappierJumpingConclusionsVisual2024} as a basis for rigor in our methodology. With this, we aim to explore the following research questions:

\begin{itemize}[leftmargin=2.75em, labelsep=0.5em]
    \item[\bf RQ1:] What are the shared design patterns and classes of visual output from software testing tools?
    \item[\bf RQ2:] How are various testing-related metrics displayed or visualized in this visual output?
\end{itemize}

Through this work, we visually examined the output of 50 software testing tools and harnesses (44 with CLI output, 6 with GUI output) across four popular programming languages, discovered shared visual elements found in outputs, cross-compared their visual layout, and explored how common interface elements (status details and test suite summaries) display and visualize important information. Drawing from these findings, we discuss the use of colour and GUIs in testing outputs to convey information and identify visual trends for developers of testing harnesses.

\section{Related Work}

To our knowledge, no other work exists in the visual analysis of software testing tool output. As such, we explore existing work at the intersection of HCI and software testing and the use of visual methodologies within HCI to ground our work.

\subsection{HCI and Software Testing}

\begin{figure*}[!ht]
    \centering
    \includegraphics[width=1\textwidth]{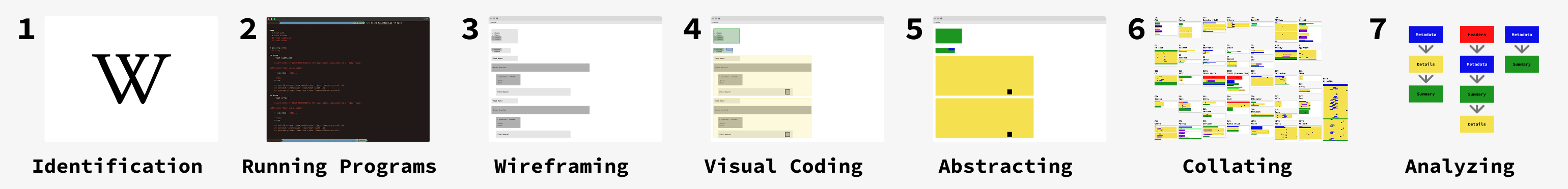}
    \caption{Our visual comparative analysis methodology phases, adapted from Frappier et al.~\cite{frappierJumpingConclusionsVisual2024}.}
    \Description{The visual comparative analysis methodology shown visually. The stages are presented sequentially from left to right: (1) Identification, (2) Running Programs, (3) Wireframing, (4) Visual Coding, (5) Abstracting, (6) Collating, and (7) Analyzing.}
    \label{fig:method}
\end{figure*} % this is here for layout purposes

Most prior work related to software testing and HCI has focused on GUI testing tools, and various platforms have been created to support playback or automation. Alaboudi and LaToza's {\it Hypothesizer}~\cite{alaboudi_hypothesizer_2023} enables hypothesis-driven playback testing to aid a developer in determining the cause of a bug in a web-based system. Chen et al.~\cite{chen_improving_2020} augment crowdsourcing-driven GUI testing by showing test traces visually in real time. For Android devices, research has explored both the augmentation of manual GUI testing through hints~\cite{liu_guided_2022} and automatic testing through the use of UI patterns as heuristics~\cite{morgado_impact_2019}. Where these studies are primarily in creating novel tools to aid users in testing GUIs, they show the importance of visual cues and outputs in the software testing lifecycle.

Additional work in software testing has adapted methods and techniques from HCI to better understand the perspective of software testers as a user group. For instance, \citet{evansBreakingTesterStereotypes2024} asked the question \textit{``Who is Testing?''} and investigated the backgrounds of over 70 industry testers to learn about differences in needs and problem-solving preferences. \citet{borgCarmenBetterGeorge2015} propose using methods from HCI to learn about the various testing techniques used and what types of bugs are found by testers with and without formal training. These works show a need for understanding of what testing tools output from a visual perspective such that future testing tools and output can be designed with these considerations in mind.

\subsection{Visual Methodologies in HCI}

Though visual output is often analyzed in qualitative HCI and visualization work, methods to work with visual data are underdeveloped, requiring the adaptation of existing qualitative methods for textual data~\cite{wangSummaryWorkshopVisual2024}. In the social sciences, visual research is more established, with photos and videos being integrated in ethnographic research through analysis of existing photos, {\it photo-elicitation} studies, or having participants take photos as part of research~\cite{sealeVisualMethods2004}. Where prior qualitative work in HCI has used photo-elicitation as a primary methodology~\cite{dirksAmplifyingCulturalValues2025, gormParticipantDrivenPhoto2017}, visual methods are not often used for analyzing existing technological artifacts. This form of analysis may aid in understanding and exploring design patterns for future work in the design of technological systems~\cite{wangSummaryWorkshopVisual2024}.

Early work in this space includes understanding common design patterns in: data comics~\cite{bachDesignPatternsData2018}, data stories~\cite{yangDesignSpaceApplying2022}, and composite visualizations~\cite{javedExploringDesignSpace2012}. An exemplar paper in the use of visual methodologies within HCI research is Frappier et al.~\cite{frappierJumpingConclusionsVisual2024}'s visual comparative analysis of online debate platforms. In their work, they perform a 7-phase visual analysis process to compare features, understand interface patterns, and discuss design implications for future work and development in debate platforms. Our work builds upon this prior work in visual methods, adapting prior methodologies to understand the design patterns behind visual output in software testing systems.

\section{Methodology}

To ground our analysis, we adapt our visual comparative analysis from Frappier et al.~\cite{frappierJumpingConclusionsVisual2024}. We outline each phase of our study as such (visualized in \autoref{fig:method}):

{\bf Phase 1.} We first identified a list of testing frameworks to use as a basis for our analysis. Adapting Wikipedia's {\it List of unit testing frameworks}~\cite{ListUnitTesting2026}, we selected a subset of popular languages (C, .NET, JavaScript, and Python) and collated each list in a spreadsheet.

{\bf Phase 2.} We then systematically ran each program. Programs were excluded from our analysis if they were deprecated, required payment to use, did not work with current systems, or if they required more than 10 minutes of overhead time to run successfully. To test each testing harness, we created a set of simple ``calculator'' functions expecting numeric or integer types: fault-free {\tt add()} and {\tt divide()} functions, and a faulty {\tt subtract()} function which adds two numbers together. Our test harnesses were then created with 2 passing tests ({\tt add()} and {\tt divide()}) and 2 failing tests ({\tt subtract()} and {\tt add()} with non-numeric input). When possible, we ran each program using a verbose mode or flag.

159 programs were tested across all selected languages and 50 (44 CLI and 6 GUI) were included in our final analysis. The final list of included programs can be found our appendix (\autoref{tab:program_table}).

{\bf Phase 3.} We then created ``reverse wireframes'' of each program. All included outputs were screenshotted and added to a Figma\footnote{\url{https://www.figma.com/}} Design file. The second author, with a background in digital design, created a set of wireframe components inductively. As many CLI outputs were textual rather than blocks, similar text was abstracted into blocks to have a unified style (e.g., large stack traces were represented by a block with the text ``{\tt <Stack Trace>}'' within). GUI outputs were kept more true-to-form, akin to an actual wireframe.

{\bf Phase 4.} We then visually coded the wireframes. As a group, we inductively created an initial set of codes on 3 wireframes. To validate our understanding of these initial codes, we coded a small subset (9) separately and met to resolve any discrepancies. The remainder were split evenly among all authors, and all authors met when coding was complete to ensure agreement. We met often to discuss codes as stated in recommendations for rigor in visual studies~\cite{meyerCriteriaRigorVisualization2019}. In line with work surrounding norms related to inter-rater reliability in HCI work, where our approach is {\it reflexive}, it is not necessary to report any inter-rater reliability metric~\cite{mcdonaldReliabilityInterraterReliability2019}.

{\bf Phase 5 and 6.} We then abstracted our coded wireframes, merging overlapping codes into a singular colour. The abstract wireframes were assembled into a large mosaic to facilitate a comparative analysis between outputs.

{\bf Phase 7.} We lastly analyzed our visual data collaboratively. Analysis surrounding visual hierarchy was facilitated by the mosaic, further abstracting the screenshots by ignoring nested codes. Other analyses returned to the wireframes and screenshots in order to explore in further detail the common interface elements and types of visualizations used within test status details and test summaries.

\section{Results}

\begin{figure*}[!th]
    \centering
    \includegraphics[width=1\textwidth]{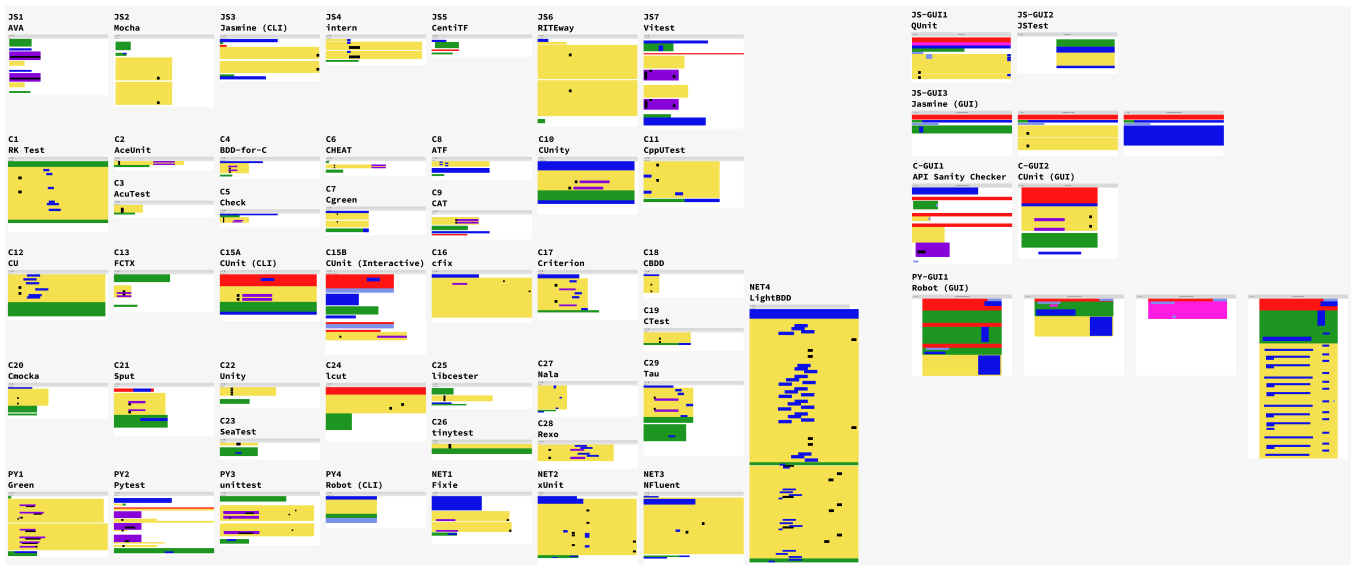}
    \caption{The full mosaic of all testing outputs, each composed of the 8 common interface elements. The CLIs and GUIs are grouped, with the former on the left, and the latter on the right. Multi-page GUIs (i.e., Jasmine GUI and Robot GUI) are displayed side by side.}
    \Description{The full mosaic of all testing outputs, each composed of the 8 common interface elements. The CLIs and GUIs are grouped, with the former on the left, and the latter on the right. Multi-page GUIs (i.e., Jasmine GUI and Robot GUI) are displayed side by side.}
    \label{fig:mosaic}
\end{figure*}

Drawing from our visual analysis, we were able to understand common interface elements and design patterns with respect to the visual hierarchy (RQ1), and how test statuses and summaries are displayed in detail in harness outputs (RQ2).

\subsection{Common Interface Elements}

\begin{figure}[ht]
    \centering
    \includegraphics[width=1\linewidth]{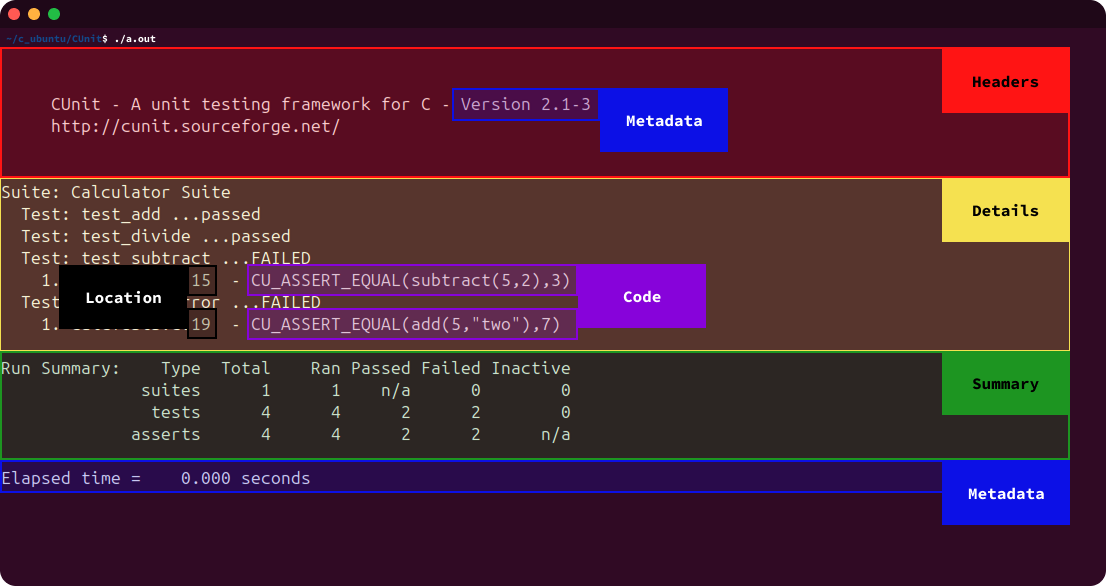}
    \caption{A screenshot of CUnit CLI with interface element codes from the coding process overlaid.}
    \Description{A screenshot of CUnit CLI with interface element codes, represented as transparent coloured boxes from the coding process overlaid. A "Headers" box is over the CUNIT header, with a smaller "Metadata" box over the version number. After, a "Details" box is over the test suite details, with smaller "Location" and "Code" boxes over the line numbers and in-line code snippets. After, a "Summary" box is over a table summary of test results, and a "Metadata" box is over the elapsed time.}
    \label{fig:cli_ex}
\end{figure}

\begin{figure}[h]
    \centering
    \includegraphics[width=1\linewidth]{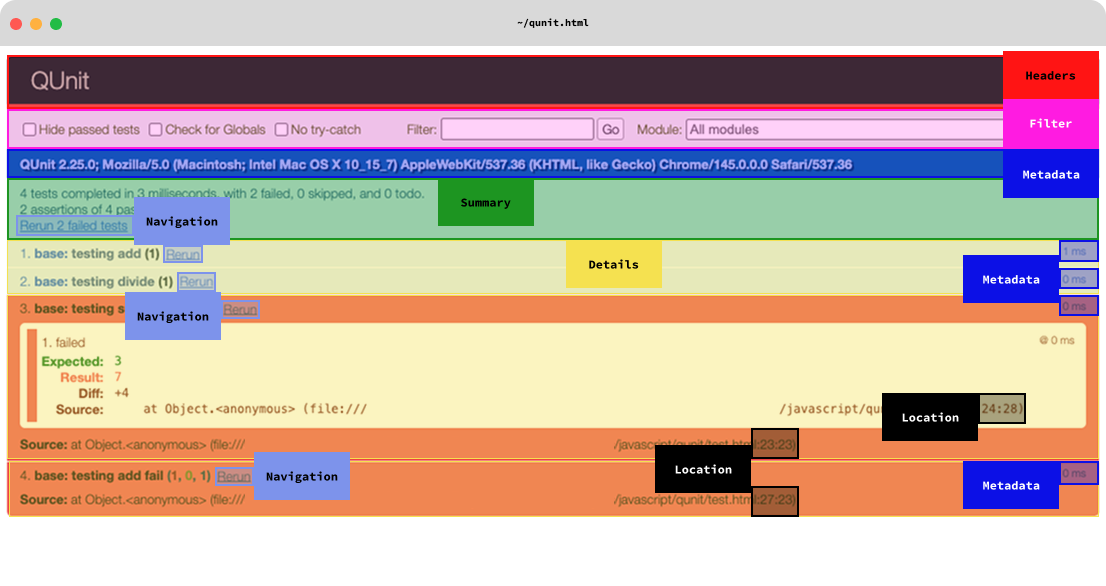}
    \caption{A screenshot of QUnit's GUI with interface element codes from the coding process overlaid.}
    \label{fig:gui_ex}
    \Description{A screenshot of QUnit's GUI with interface element codes, represented as transparent coloured boxes from the coding process overlaid. A "Headers" box is over the QUnit header. After, a "Filter" box is over checkboxes, a dropdown, and searchbar. After, a "Metadata" box is over the system information. After, a "Summary" box is over a textual summary of test results, and a "Navigation" box is over a button to rerun tests. After, a "Details" box is over the test suite details, with smaller "Location" boxes over the line numbers, "Metadata" over the timing, and a "Navigation" box is over buttons to rerun tests.}
\end{figure}

From our coding process, we discovered 8 common interface elements within the 50 test harness outputs. Concrete examples of these elements are seen in \autoref{fig:cli_ex} (a CLI) and \autoref{fig:gui_ex} (a GUI).

\subsubsection{Test Status Details} Test status details (\colorbox{myyellow}{\tt Details}) are sections of output that contain detailed information about the test(s) that was/were conducted. These details include, but are not limited to: stack traces, code lines, error location indicators, and reasons for failure.

\subsubsection{Test Summary} Test summaries (\colorbox{mygreen}{\tt Summary}) display the results of a test suite execution with reduced granularity and are often more succinct compared to test status details. They focus on showing the overall results of the test suite execution at a glance rather than the particular details of each test.  

\subsubsection{Metadata} Metadata (\colorbox{myblue}{\color{white}{\tt Metadata}}) displays general information related to the tests or test suites. This includes, but is not limited to: details about the system (e.g., OS version), time spent per test or suite, suite name, and number of tests with no result details.

\subsubsection{Code} Code (\colorbox{mypurple}{\color{white}{\tt Code}}) elements are instances of actual code displayed in the test suite output. They either consisted of individual lines of code or multi-line blocks of code. 

\subsubsection{Error Location Identifiers} Error location identifiers \\(\colorbox{myblack}{\color{white}{\tt Location}}) are interface elements that point to the location of code failure in some manner. This includes line and character numbers as well as carats pointing to the failing location. %\td{I feel like this may be too much? More details should be put in 4.3 about them? Or should we fully describe here?}

\subsubsection{Headers} Headers (\colorbox{myred}{\tt Headers}) are generic elements either displaying the test harness name or section title. Oftentime, these headers contain brief metadata (e.g., version numbers) within.

\subsubsection{Filters} Filters (\colorbox{mypink}{\color{white}{\tt Filter}}) are any element that enables the filtering or searching of test cases within the test suite on the interface. In GUIs, these may look like checkboxes, dropdown menus, or searchbars.

\subsubsection{Navigation} Navigation (\colorbox{mylblue}{\tt Navigation}) elements are any form of tab or link used to navigate between features of a tool or refresh/rerun individual tests or full test suites.

\subsection{Visual Hierarchy}

To observe shared interface patterns among outputs, we created a mosaic of all abstract wireframes (\autoref{fig:mosaic}). In order to understand broader classes of visual hierarchies across outputs, we classified wireframes based on their large blocks of content and ignored nested or inline elements. For example, if a test status details block contains location identifiers and metadata, they are ignored and treated as part of the details block. We report our findings for CLIs and GUIs separately.

\subsubsection{CLIs} 
\begin{figure}[!h]
    \centering
    \includegraphics[width=1\linewidth]{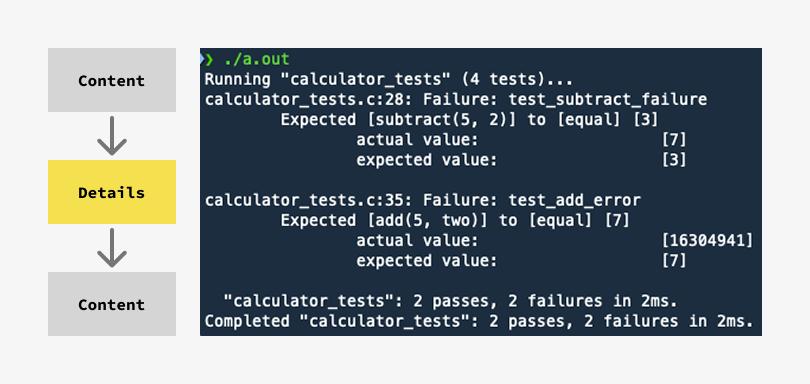}
    \caption{An example of the ``details-in-the-middle'' class of output.}
    \label{fig:detailsmiddle}
    \Description{An example of the ``details-in-the-middle'' class of output, with a flowchart of "Content" to "Details" to "Content" on the side.}
\end{figure}

From our CLIs, we found two major classes of visual hierarchies. The first, ``details-in-the-middle'' was found in 25/44 of the outputs, where a test status details block was sandwiched between other content blocks. An example of this is seen in CGreen (\autoref{fig:detailsmiddle}), where \colorbox{myblue}{\color{white}{\tt Metadata}} precedes the test status \colorbox{myyellow}{\tt Details} block, and a \colorbox{mygreen}{\tt Summary} is output last.

\begin{figure}[!h]
    \centering
    \includegraphics[width=1\linewidth]{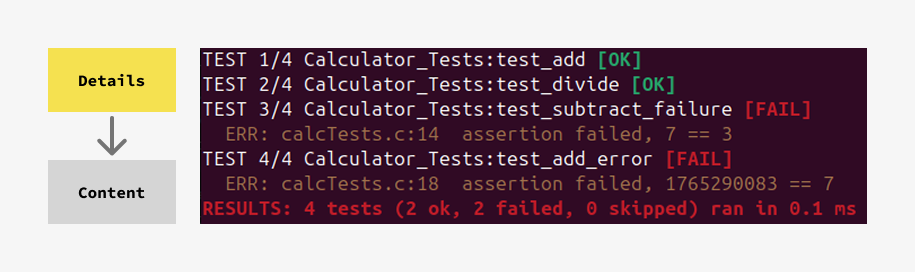}
    \caption{An example of the ``details-on-the-outside'' class of output.}
    \label{fig:detailsoutside}
    \Description{An example of the ``details-on-the-outside'' class of output, with a flowchart of "Details" to "Content" on the side.}
\end{figure}

The second most common class was ``details-on-the-outside'', found in 16/44 of the outputs. In this case, the test status details either precedes other content blocks (in 12 outputs), or follows up from a content block (in 4 outputs). An example of this is seen in CTest (\autoref{fig:detailsoutside}), where a test status \colorbox{myyellow}{\tt Details} block is output before a \colorbox{mygreen}{\tt Summary}.

\begin{figure}[!h]
    \centering
    \includegraphics[width=1\linewidth]{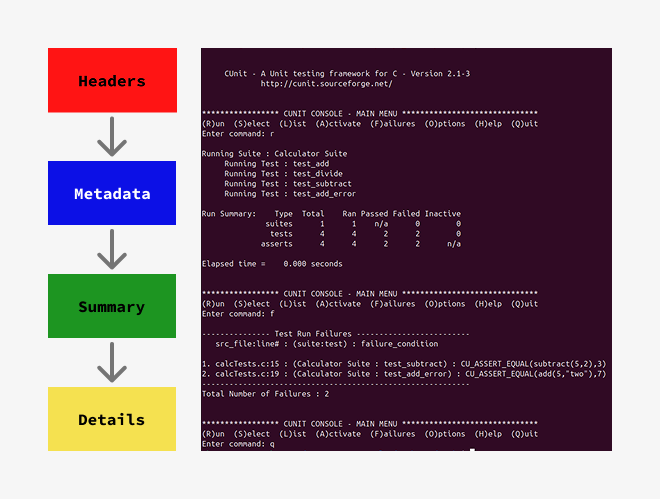}
    \caption{CUnit's interactive CLI mode, with an example visual hierarchy resulting from running the tests, displaying failures, and quitting.}
    \label{fig:cunit}
    \Description{An example of CUnit's interactive CLI mode, with an example visual hierarchy, with a flowchart of "Headers" to "Metadata" to "Summary" to "Details" on the side.}
\end{figure}

There were 3 unique CLI outputs that did not fit nicely into either class. CentiTF did not contain any test status details, going from brief metadata to a test suite summary. Rexo only consisted of one large test status details block, not including any broad summary or metadata about the test suite. Last, CUnit included an {\it interactive} CLI mode (\autoref{fig:cunit}), enabling users to navigate between running tests, viewing the failures as a report, setting test options, and quitting through single-key commands. This was the only fully interactive CLI interface we discovered in our analysis, with other systems (e.g., Vitest) having limited interactivity through features like hot-reload.

\subsubsection{GUIs} Most of our GUIs followed a similar visual pattern to our CLIs, either ``details-in-the-middle'' (CUnit GUI and JSTest), or ``details-on-the-outside'' (QUnit and API Sanity Checker). However, GUI output is able to easily include navigation, where while the 4/6 aformentioned were single-page, 2/6 GUIs (Jasmine GUI and Robot GUI) were multi-page interfaces. Of note, Jasmine GUI broke from the prior classes of outputs and displayed a test suite \colorbox{mygreen}{\tt Summary}, test status \colorbox{myyellow}{\tt Details}, and test \colorbox{myblue}{\color{white}{\tt Metadata}} across three separate tabs (named {\it Spec List}, {\it Failures}, and {\it Performance} respectively).

\subsection{Test Status Details}

% Test status (\colorbox{myyellow}{\tt Details}) are sections of output that contain detailed information about the test(s) that was/were conducted. These details include, but are not limited to, stack traces, code lines, error location indicators, and reasons for failure.

\begin{figure*}[!t]
    \centering
    \includegraphics[width=1\textwidth]{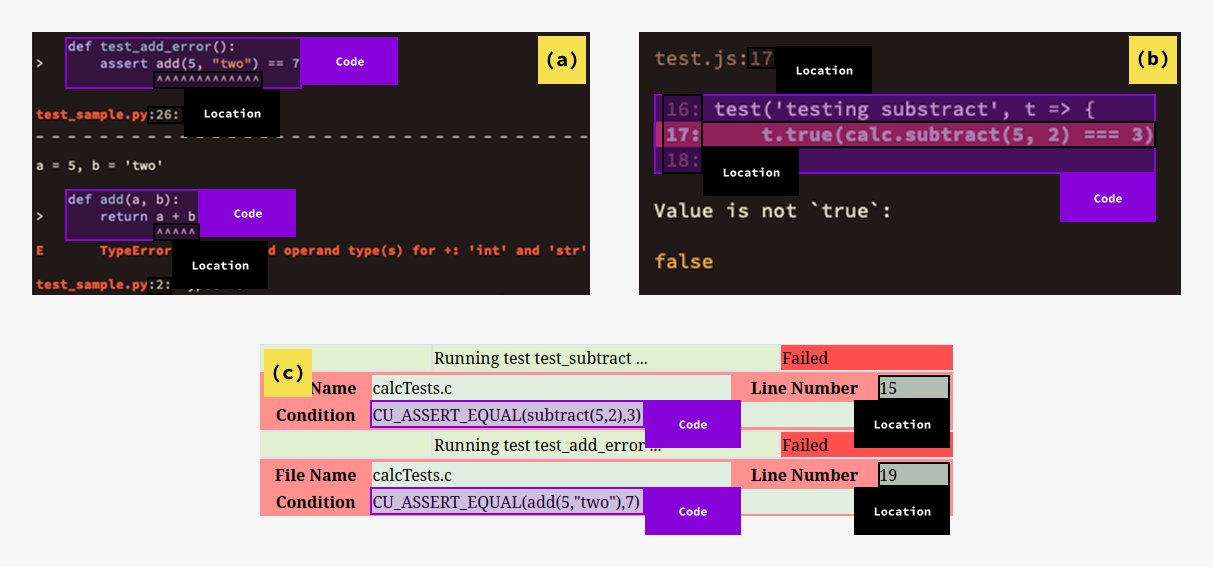}
    \caption{Examples of error location identifiers and code lines/blocks: (a) Pytest displaying line numbers, carats, and code block; (b) AVA exhibiting line number, code block, and highlight of the failing call; (c) CUnit showing line number and code line}
    \label{fig:error_id}
    \Description{Examples of error location identifiers and code lines/blocks. Over each screenshot, transparent boxes saying "Code" and "Location" are overlayed to point out code blocks or lines for the former, or carats or line numbers for the latter.}
\end{figure*}

Test status \colorbox{myyellow}{\tt Details} appear in 43/44 of the CLIs and all 6 of the GUI outputs analyzed. These sections contain all of the detailed information about the test(s) conducted and take up the majority of the output as seen in \autoref{fig:mosaic}. They contain a wide range of different types of content and formatting to display the results of a test run. This content can include: the test result or status, an error message or reason for a failure, a stack trace of the failure, section dividers, timing metadata, code lines or blocks, and error location identifiers.

% \subsubsection{Contents}

% \td{Variety of content types were found in test status details including:}
% \td{Outline numbers of each and how they're used -- small table for them?}
% \td{Provide examples}

\subsubsection{Error Identification}

Identification of an error and the details causing it is an important aspect of software testing. We found four primary ways in which error information was communicated to the tester: identification of the failing code, the location of the failing code, stack traces, and an error message or reason for the failure. 

40 of the CLIs and 4 GUIs provided some manner in which to identify the location in which an error occurred. In 43 of these cases, the tester was provided with a line number. In 2 cases, the failing function call was marked in red, with 1 program providing both (\autoref{fig:error_id}). 5 of the CLIs used carats as a supplementary identifier to point out the error location alongside line numbers (\autoref{fig:error_id}a).

Showing the failing code line or block was also a common trait in many of the outputs; 21 CLIs and 2 GUIs showed either a line or block of code (19 line, 4 block) (\autoref{fig:error_id}). Additionally, 8 outputs displayed a stack trace in order to provide a more in-depth view of the error's source and why it occurred. 

\subsubsection{Verbosity}

\begin{figure}[h]
    \centering
    \includegraphics[width=1\linewidth]{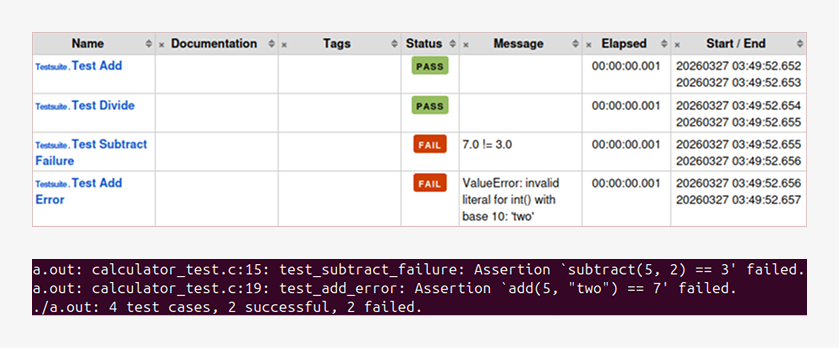}
    \caption{Example showing a differing amount of detail between outputs: (top) Robot Framework displaying a formatted table with each test case executed and its result; (bottom) AceUnit displaying only the failing test cases}
    \Description{Two test suite summary blocks from the analyzed test suite software. One shows Robot Framework, a Python testing framework, which displays a formatted table with each test case executed and its result. The second shows AceUnit, a C testing framework, which only displays the failing test cases.}
    \label{fig:verbosity}
\end{figure}

The verbosity and amount of detail in the outputs drastically varied between the testing outputs, with platforms such as AceUnit identifying the failing case in one line without any other details, and others like NFluent showing a detailed stack trace on why the test failed. Only 30 outputs (4 of which were GUIs) showed the completion status of all test cases executed. For instance, Robot Framework provides details around all of the passing and failing test cases, including the reasoning behind those failures, while AceUnit only displays the failing test cases (\autoref{fig:verbosity}).

Within test status details, 13 CLI and 2 GUI outputs included some \colorbox{myblue}{\color{white}{\tt Metadata}}. The only metadata found within test status details were values displaying how long a test case ran and/or how long a test suite took to complete. Some of the other content included \colorbox{myred}{\tt Headers} or section dividers (9 CLIs, 5 GUIs) to visually split up and organize content, or methods of \colorbox{mypink}{\color{white}{\tt Filtering}} or \colorbox{mylblue}{\tt Navigating} to ``minimize'' or change content, reducing the amount of information on the screen at once (1 CLI, 4 GUIs).

\subsection{Test Summaries}

In our analysis, differentiating test \colorbox{mygreen}{\tt Summary} and test status\\ \colorbox{myyellow}{\tt Details} was difficult to do as they appeared to overlap in both function and content within some of the test tools. After individually coding a number of test interfaces, researchers met and compared the definitions of each code to determine how to differentiate them going forward. Researchers concluded that test summaries were composed of a high-level overview of all tests run, and contained no details about the individual tests themselves. Test status details contain specific information about the test that was run, and should contain only one test's information within it. 

\begin{figure*}
    \centering
    \includegraphics[width=1\textwidth]{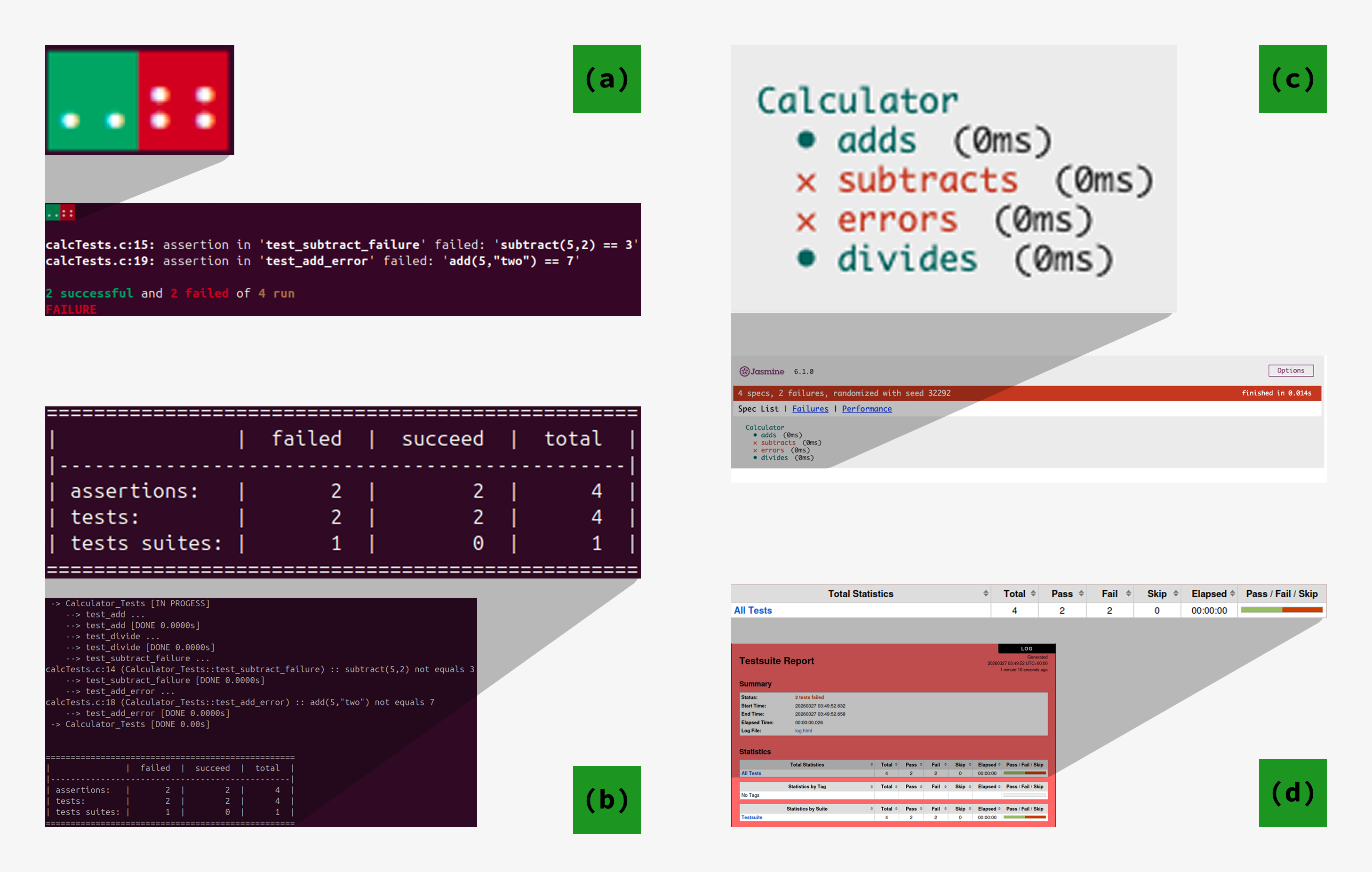}
    \caption{Example test suite summary blocks from our sample: (a) CHEAT, using two colours with period and colon symbols to create a basic summary visualization; (b) CU, using ASCII characters to create a table, a complex summary visualization; (c) Jasmine's GUI, using different list bullet shapes and colours to create a basic summary visualization; and (d) Robot Framework's GUI, using a table and a red and green rectangular graphic to create a complex summary visualization}
    \Description{Four test suite summary blocks from the analyzed test suite software. A. CHEAT a C software which uses two colours red and green, with a period and colon symbols to denote pass and fail which creates a basic summary visualization.}
    \label{fig:summaries}
\end{figure*}

Within most test \colorbox{mygreen}{\tt Summary} sections, we observed frequent usage of \colorbox{myblue}{\color{white}{\tt Metadata}} (in 37/44 CLIs) to illustrate additional information about the tests. Information such as amount of time taken, system, file, and test harness information, and stack traces were common within the test summary sections.

The summaries themselves were also presented in a variety of formats to help visualize the results. We defined both basic (27/44) and complex visualizations (17/44) within a test summary. Basic visualizations were often not the main focus within the interface, instead acting as a small addition to quickly ascertain the results of a test harness. CHEAT, a C test framework, used a combination of red and green periods and colons to define whether a test case had passed or not (\autoref{fig:summaries}a). Complex visualizations were often a major focus within the interface, being integrated alongside other summary components. These visualizations expressed more complex ideas and did not always limit themselves to binary pass and fail results from a test harness. The C test framework CU highlights this with a complex ASCII table highlighting passed and failed assertions, tests and test suites (\autoref{fig:summaries}b).

% \begin{figure}[!h]
%     \centering
%     \includegraphics[width=1\linewidth]{figures/basicvis_cli.png}
%     %\includegraphics[width=1\linewidth]{figures/CCHEAT.pdf}
%     \caption{A C test framework CHEAT, using two colours with the period and colon symbols to create a basic summary visualization.}
%     \label{fig:CCHEAT}
% \end{figure}

% \begin{figure}[!h]
%     \centering
%     \includegraphics[width=1\linewidth]{figures/complexvis_cli.png}
%     %\includegraphics[width=1\linewidth]{figures/CCU.pdf}
%     \caption{A C test framework CU, using ASCII characters to create a table, a complex summary visualization.}
%     \label{fig:CCU}
% \end{figure}

Summaries within GUIs were categorized in a similar fashion, with basic summaries being contained within advanced UI elements such as the JS framework Jasmine (GUI mode) (\autoref{fig:summaries}c). More complex summaries were also present, not only utilizing more advanced UI elements but also having more comprehensive visual elements such as bar graphs and containing links to test status details such as Robot Framework's GUI mode for Python (\autoref{fig:summaries}d).

% \begin{figure}[!h]
%     \centering
%     \includegraphics[width=1\linewidth]{figures/basicvis_gui.png}
%     %\includegraphics[width=1\linewidth]{figures/JSJasmine.pdf}
%     \caption{A JS test framework Jasmine, using different list bullet shapes and colours to create a basic summary visualization.}
%     \label{fig:JSJasmine}
% \end{figure}

% \begin{figure}[!h]
%     \centering
%     \includegraphics[width=1\linewidth]{figures/complexvis_gui.png}
%     %\includegraphics[width=1\linewidth]{figures/PythonROBOT.pdf}
%     \caption{A Python test framework ROBOT, using a table and a red and green rectangular graphic to create a complex summary visualization.}
%     \label{fig:PythonROBOT}
% \end{figure}

\subsection{Use of Colour}

Throughout both CLI and GUI test harnesses, the testing tools utilized colour in a variety of different ways. Within CLI test harnesses, we observed four categories of colour usage: no colour (15/44), one colour (4/44), two colours (13/44), and three or more colours (12/44). Within GUI interfaces, we observed only two colours (2/6) or three or more colours (4/6) being used. Colours used within the test harnesses varied with the exception of two: red was always used to indicate a failure state for a test or test suite, and green was always used to indicate a success state for a test or test suite.

\section{Discussion}
In line with previous qualitative work in software testing~\cite{strandberg_information_2019}, our analysis found that the use of colour and the GUI outputs in software testing programs both have implications across software testing development and accessibility. We discuss this range of visualizations from colour to GUIs in supporting software testing practices.

\subsection{Implication of Colour}

From our analysis, we uncovered basic trends regarding the usage of colour to inform future testing output interface designs. In CLIs, a surprising number of programs did not utilize colour. Where some used ASCII constructions (e.g., tables) to highlight metrics in place of colour, some interfaces were purely textual, containing no distinct visual elements. In these cases, the user is required to parse through lines of information without visual aids, akin to that of a dense stack trace. 

Of the four CLIs that used only one colour, all chose red. While this allowed the test program to easily draw attention to failures, the application of this colour varied greatly. In one example the interface excessively used it, highlighting the lines containing the test case, an error message, a file name, the line numbers and expected and actual test values. This made it difficult to parse out specific information. In another example it was underused, with only an error message being highlighted red. Instead, this interface attempted to make use of spacing with the remaining uncoloured text to try and draw emphasis to important points of failure. Of the 13 CLI programs that used two colours, each used green and red to highlight passing and failing tests. Many of these interfaces still utilized advanced ASCII decoration and layouts to help divide information but were successfully able to utilize these colours to contrast the success and fail states. The utilization of three or more colours in 12 of the CLI interfaces was the most interesting. In addition to utilizing green and red, test programs included a variety of colour choices such as blue or brown to highlight code, indicate running tests or act as additional dividers. Some of the test harnesses used a third colour to highlight metadata such as the version of the language compiling the program. Interestingly, despite the choice to use multiple colours, we found that some interfaces still underutilized them, choosing to only highlight keywords such as ``Failed'' and ``Passed'', leaving blocks of uncoloured text for the user to parse through. 

In the GUI designs, colour was also used in a similar fashion, focusing on red and green to denote pass and fail, and adding in additional colours to highlight other information. Due to the visual nature of GUIs colour was applied in more interesting ways, such as changing the colour of tables or the background to green or red depending on the test suite results. Some of the GUI interfaces excessively applied colour making it difficult to parse them quickly when looking for key information. Unlike the CLI tools, GUIs did not limit the usage of green and red to a success or fail state. One of the GUIs utilized green for arrows pointing to test case names whose font colour would change based on failure. Another GUI used three different green tones for the background colour of two separate tables and for a success state.

Throughout all the test programs the choice of colours can also be questioned from an accessibility angle. Red-Green colour blindness would make many of these interfaces difficult to parse countering the goal of making the results stand out. Additionally, the lack of consistency in the colour applications can also help or hinder a user attempting to read the results. In the GUIs, we found a number of potential issues based on previous research in colour blindness. The choice of colour palettes, the usage of overly contrasting colours, and the reliance on the colour red all contribute to a less accessible interface \cite{katsnelsonColourMeBetter2021}.

\subsection{GUI-based Outputs in Software Testing}

In our sample set of 50 testing tool and harness outputs, only 6 were GUIs. Although difficult to generalize the GUI results given our small sample, they did demonstrate some interesting trends that are worth noting and investigating further.

Many GUIs split their output across multiple tabs or pages. Previous work exploring pagination or the sectioning of information shows that it has an impact on cognitive load~\cite{ayresSplitAttentionPrincipleMultimedia2005, hollenderIntegratingCognitiveLoad2010}. This work outlines the split-attention principle, which states that when information is split across a variety of disparate sources, the cognitive load required to fully understand the information increases. As such, we observe that the GUI outputs in our dataset could be subject to this principle if key information required to understand a test case is split between interface elements.

When looking at the mosaic in \autoref{fig:mosaic}, it shows that the formatting of the GUI outputs are much less dense than the CLI outputs. Test status details take up a larger amount of screen space in the CLI outputs than they do on the GUI outputs. The reduced density of the GUIs due to styled HTML or XML formatting can aid in making the information being communicated easier to visually parse.

The only GUIs which displayed failing code were made for the C language. How the code was displayed in both of these cases was relatively abstract compared to the usage in some of the CLIs. In one of the two instances, it showed a block of code made by the harness to run the failing test and highlighted the failing function call in red text. The second instance showed the assert line that failed alongside the file name and line number. Since the use of code was only in C testing tool GUIs and not in GUI outputs for more modern languages, it may indicate that the current focus is more on the reason for failure and where the tester can find the failing code to reduce the amount of info displayed, however, this would require more study to form a conclusion.  

4 of the GUI-based outputs and 1 CLI output had interactivity as a feature. This interactivity took multiple forms from navigating between different tabs or pages of information to the closing or opening of accordion menus to selectively view specific information. The inclusion of interactivity in most GUI outputs is interesting as it allows for the tester to customize the information that is being presented to them at any given time rather than being shown one static output from a test suite execution on a CLI. These interaction methods and elements support the \textit{Overview}, \textit{Filter}, and \textit{Details-on-demand} tasks outlined by \citet{shneidermanEyesHaveIt1996}, which allow testers to reduce the amount of visible information to a manageable amount and avoid becoming overwhelmed.

\section{Limitations}

This paper presents a visual analysis of a small subset of software testing platforms and harnesses. In our searching, apart from the Wikipedia list~\cite{ListUnitTesting2026}, there were no other clear repositories of software testing tools across multiple languages. As such, we recognize that the list is not exhaustive or up-to-date; many links and tools were inaccessible or led to deprecated projects. Future work should incorporate both more modern tooling frameworks for software testing and tools beyond our subset of languages (e.g., Java).

We also recognize that the program adapted for each test harness is simplistic and not realistic compared to modern unit testing practices. This was an intentional trade-off, where our goal was to examine a breadth of programs across multiple languages rather than achieve full ecological validity for a smaller subset of frameworks. Future work may explore the outputs of various test frameworks using real-world test harnesses to compare more realistic outputs.

% \td{Just copying over limitation and future work points from presentation:}
% \begin{enumerate}
%     \item Limitations:
%     \begin{enumerate}
%         \item Wikipedia list is not exhaustive or up-to-date -- Many links were dead, led to inaccessible Google code archives, or deprecated projects
%         \item Limited analysis on other popular languages (e.g., Java) -- Result of time constraint for course project
%         \item Test harness used had limited test in order to view output -- Tradeoff between ease-of-translation between languages/harnesses and ecological validity
%     \end{enumerate}
%     \item Future Work:
%     \begin{enumerate}
%         \item Informs the design of GUI test harnesses from common patterns found in CLIs in order to create a seamless experience between the two
%         \item User study to explore usability and cognitive load of testing platforms’ visual output in mind -- Explore preferences between CLIs following “Details-in-the-Middle” vs. “Details-on-the-Outside” -- Explore pagination vs 1-page options in GUIs
%         \item Expand on visual analysis to include other popular languages (e.g., Java)
%     \end{enumerate}
% \end{enumerate}

\section{Conclusion}

In this work, we conducted a visual comparative analysis of 50 software testing tools and harnesses (44 CLIs and 6 GUIs) across C, .NET, JavaScript, and Python. Overall, we found 8 common visual interface elements, 2 broad classes of visual interface patterns, and explored how metrics, colour, and visualizations are used within test summaries and test status details. Developers of testing tools and harnesses can use our results to inform their design of both CLI and GUI outputs to create a seamless experience between each. Future HCI or empirical software testing work exploring the visual output of software testing tools may use these design patterns and classes in a user study exploring usability or user preferences between ``details-in-the-middle'' and ``details-on-the-outside'' interfaces, or on pagination versus scrolling in GUIs. Beyond software testing, we contribute additional work using visual methods in empirical HCI research and work towards a robust, rigorous, and reflexive methodology for working with visual data.

% --------------------------------

%% Acknowledgements 
%% ASK DAN WHICH ONES TO USE:
% \begin{acks}
% This work was made possible by 
% NSERC Discovery Grant 2024-03827,  
% Canada Foundation for Innovation Infrastructure Fund 33151 ``Facility for Fully Interactive Physio-digital Spaces,''  
% NSERC USRA Program for undergraduate research, 
% Waterloo-Inria INPUT Associate Team
% \end{acks}

%% reference section
\bibliographystyle{lib-acm/ACM-Reference-Format}
\bibliography{_references.bib}

% SUBMITTING TO ARXIV
% If you are submitting your paper to arXiv, change the bibliography to be "main_acm.bib" (the same name as your main.tex file) and use the Submit feature in OL to compile the main_acm.bbl file (select arXiv as the submission type). To avoid processing errors on arXiv with your references, your bibliography needs to have the same name as your main.tex file before you compile it.

% UPDATE: I think the new ACM format messes up the compilation: you now have to add the lib files because they aren't automatically downloaded now. And ArXiv now seems to prefer .bib, so try renaming the .bib file and just downloading the source as-is. Haven't tried it myself but might be fine now.

% Use the arxiv latex cleaner (https://github.com/google-research/arxiv-latex-cleaner) to clean the source before you upload it to arXiv. This is also useful when you upload the source to PCS:
%  arxiv_latex_cleaner <folder> --keep_bib --commands_to_delete <commands, e.g., dv> 
%  You can add "--commands_only_to_delete rev" to your archix latex cleaner command to delete rev commands, but keep the rev-text (unlike --commands_to_delete which deletes the inner text too)

%% appendices
%% If your work has an appendix, this is the place to put it. The TC: comments tell the word count scripts to ignore appendix.  

%TC:ignore  
\appendix
\makeatother
\clearpage
% Reset the figure count and add an A prefix to distinguish it from your other figures
\renewcommand\thefigure{\thesection.\arabic{figure}}
\renewcommand\thetable{\thesection.\arabic{table}}
\setcounter{figure}{0}
\setcounter{table}{0}
%\cite{}
\section{Included Programs}

% appendix figures and table should always be "here" layout with [h]
\begin{table}[h]
    \caption{List of all programs included in our visual analysis}
    % link to table tex

\small % smaller table text
% \textbf{\setlength\tabcolsep{3.3pt} % default value: 6pt
% \renewcommand{\arraystretch}{1.5}}
\begin{tabular}{l|l|c|l}
\toprule
\textbf{Identifier} & \textbf{Tool Name} & \textbf{Language} & \textbf{Type of Output} \\
\midrule
JS1 & AVA & JavaScript & CLI\\
JS2 & Mocha & JavaScript & CLI\\
JS3, JS-GUI3 & Jasmine & JavaScript & CLI, GUI\\
JS4 & intern & JavaScript & CLI\\
JS5 & CentiTF & JavaScript & CLI\\
JS6 & RITEWay & JavaScript & CLI\\
JS7 & Vitest & JavaScript & CLI\\
JS-GUI1 & QUnit & JavaScript & GUI\\
JS-GUI2 & JSTest & JavaScript & GUI\\
\midrule
C1 & RK Test & C & CLI\\
C2 & AceUnit & C & CLI\\
C3 & AcuTest & C & CLI\\
C4 & BDD-for-C & C & CLI\\
C5 & Check & C & CLI\\
C6 & CHEAT & C & CLI\\
C7 & Cgreen & C & CLI\\
C8 & ATF & C & CLI\\
C9 & CAT & C & CLI\\
C10 & CUnity & C & CLI\\
C11 & CppUTest & C & CLI\\
C12 & CU & C & CLI\\
C13 & FCTX & C & CLI\\
C15, C-GUI2 & CUnit & C & CLI, Interactive (CLI), GUI\\
C16 & cfix & C & CLI\\
C17 & Criterion & C & CLI\\
C18 & CBDD & C & CLI\\
C19 & CTest & C & CLI\\
C20 & Cmocka & C & CLI\\
C21 & Sput & C & CLI\\
C22 & Unity & C & CLI\\
C23 & SeaTest & C & CLI\\
C24 & lcut & C & CLI\\
C25 & libcester & C & CLI\\
C26 & tinytest & C & CLI\\
C27 & Nala & C & CLI\\
C28 & Rexo & C & CLI\\
C29 & Tau & C & CLI\\
C-GUI1 & API Sanity Checker & C & GUI\\
\midrule
PY1 & Green & Python & CLI\\
PY2 & Pytest & Python & CLI\\
PY3 & unittest & Python & CLI\\
PY4, PY-GUI1 & Robot Framework & Python & CLI, GUI\\
\midrule
NET1 & Fixie & .NET & CLI\\
NET2 & xUnit & .NET & CLI\\
NET3 & NFluent & .NET & CLI\\
NET4 & LightBDD & .NET & CLI\\
% \addlinespace[0.15cm]
\bottomrule
% \hline %%%%%%%%%%%%%%%%%%%%%%%%%%%%%%%%%%%%%%%%%%%%%%%%%%%%%%%%%%%%%%%%%%%%
\end{tabular}
    \label{tab:program_table}
\end{table}

%TC:endignore 

\end{document}